\documentstyle[aps,prb]{revtex}

\newcommand{\dm}{D^{-}} \newcommand{\dz}{D^{0}}
\newcommand{\dmn}{$D^-$} \newcommand{\dzr}{$D^0$} \newcommand {\rv}
{\vec{r}} \newcommand {\rhv}{\vec{\rho}} 

\begin{document}

\title{ Off-center \dmn {} centers in a quantum well in the presence
  of a perpendicular
magnetic field: angular momentum transitions and magnetic evaporation }

\author{C. Riva,\cite{clara}  V. A. Schweigert,\cite{vitaly} and 
F. M. Peeters\cite{francois} \ \
\\Departement Natuurkunde, Universiteit Antwerpen (UIA),
Universiteitsplein 1, B-2610 Antwerpen, Belgium  } \date{\today}
\maketitle
\begin{abstract}
We investigate the effect of the position of the donor in the quantum well
on the energy spectrum and the oscillator
strength of the \dmn {} system in the presence of a perpendicular
magnetic field. As a function of the magnetic field we find that
 when \dmn {} centers are placed sufficiently off-center they undergo
 singlet-triplet transitions which are similar 
to those found in many-electron parabolic quantum dots. The
main difference is that the number of such transitions depend on the
position of the donor and only a finite number of such singlet-triplet
transitions 
are found as function of the strength of the magnetic field.
For sufficiently large magnetic fields the two electron
system becomes unbound. For the near center \dmn {} system no
singlet-triplet and no unbinding of the \dmn {} is found with increasing
magnetic field. A magnetic field vs. donor position phase
diagram  is presented that depends on the width of the quantum well.
\end{abstract}


\twocolumn

\section{Introduction}

In multilayer and quantum well structures, such as
$GaAs/Al_xGa_{1-x}As$, the electrons bound to donor impurities
situated in the barrier 
tend to migrate in the well, due to the favorable potential gap. There
 they are trapped by
the impurity donors, such as $Si$, that are naturally or artificially
present in the material. The trapping of one electron by a
 donor does not completely
screen the charge of the donor itself,  thus  bounded states of negative
charged donors are in principle and in practice possible.\cite{Huant}

A great deal of attention has been given to the formation and stability
of negative donor  centers in semiconductors 
in recent years. Those systems, indeed, being the simplest many-body
system, represent  an
interesting occasion  to study the electron-electron interactions in solids.

In previous experimental and theoretical studies the
dependence of the binding energy of the \dmn {} on the magnetic
field strength and on the dimension of the quantum well have been done.
While a great part of these works consider the on-center \dmn {}
problem,\cite{Shi,Pao,Djub}  
i.e. when the impurity donor is at the center of the well, the study of the
off-center, i.e. when the impurity donor is displaced from the center of the
well,  and the barrier \dmn {} problem, i.e. when the donor is in the barrier,
are  much less investigated. On the theoretical side, Zhu and Xu\cite{Zhu} studied
the spin-singlet 
L=0 and
the spin-triplet L=-1 states  for a quasi-2D \dmn {}  while Fox and Larsen\cite{larsen} studied the barrier \dmn {} in which the electrons
are moving in a perfect 2D plane.
The dependence of the properties of a \dmn {} system on the position
of the donor with respect to the center has been partly 
investigated by  Marmorkos {\it et al.}\cite{Marmorkos} They considered
the problem of  
 a double quantum well in which one of the two wells hosts, in its center, the
 donor, while the other contains the electrons. On the experimental side
 we point out the work of Jiang {\em et al.}\cite{McCombe} in which
 experimental evidence of an off-center \dmn {} system was presented.
All these studies on off-center and barrier \dmn {} show spin-singlet
spin-triplet transitions of the ground state with increasing strength
of the magnetic field.
But the situation studied in previous works differs from the real
problem of the 
off-center \dmn {}. The work of Zhu {\it et al.} is most close to the
real experimental situation but they  
studied only the first two state of the \dmn {} system. Such
singlet-triplet transitions have also been 
observed in electron systems confined in quantum dots and are
known  as {\it magic magnetic number} ground state transitions. 
\cite{magic}
In quantum dots the electrons are  held
together by a parabolic or hard wall confinement  potential which for
the \dmn {} problem is replaced by the Coulomb potential of the donor
impurity.  Thus it
seems that the appearance of singlet-triplet transitions is a characteristic feature of confined electronic systems,
and in this paper we will shed more light on the condition under which
such transitions appear in the \dmn  {} system. 
The \dmn {}  problem has the added flexibility that the 
singlet-triplet transition can be influenced by changing the position of
the donor with respect to the center of the quantum well. It is  even
possible that for certain donor positions there is no singlet-triplet
transition at all. 

In the present paper we study the properties of the  
off-center \dmn {} as function of  the position of the donor in the
well, and as function of  the quantum well width in the quasi-2D
approximation. In Sec. \ref{model} we present our model and explain
how we obtain the wave function and energy of the different \dzr {}
and \dmn {} levels. Next, in Sec. \ref{enesp}, we present and discuss
the energy spectral behavior 
for    
quantum wells of width 200\AA {} and  100\AA. Then, 
 we compare the
results of the two calculations in order to have a better
understanding of the reasons that underlie  the
different  behaviors of the two energy spectra.
Next, in Sec. \ref{fo}, we evaluate and study the dependence of the oscillator
strength and of the transition energies on the magnetic field and
on the position of the donor with respect to the center of the well.
In Sec. \ref{espdata}  we use our model to explain 
 the cyclotron resonance
experiment of Jiang
{\it et al}.\cite{McCombe}
Our conclusions are presented in Sec. \ref{conclusion}
\section{The model}
\label{model}

The properties of the off-center \dmn {} in a finite-height quantum
 well under the influence of a perpendicular magnetic field will be
 treated in the present paper.
In the frame of the effective mass approximation the Hamiltonian of
 the \dmn {} system is given by
\begin{equation}
H_{{\dm}}=H^{\dz}_1+H^{\dz}_2+V_{ee}(|\rv_1-\rv_2|),
\label{dm-hamil}
\end{equation}
where $H^{\dz}_i$ is the Hamiltonian for the i-th one electron \dzr\ 
 system and $V_{ee}$ is the electron-electron repulsive Coulomb
 interaction. Using cylindrical coordinates and the effective Bohr
 radius, $a_B= \hbar^2 \epsilon_0 / m^* e^2$, and the effective Rydberg, $R_y=
 e^2/2\epsilon a_B$,  as units of length and 
 energy respectively, the neutral donor Hamiltonian $H_i^{\dz}$ and
 the electron-electron Coulomb
 potential assume the form
\begin{eqnarray}
&H^{\dz}_i=-\nabla^2+{\gamma \over i}{\partial \over \partial
\phi_i}+{1 \over 4} \gamma^2 \rho^2_i -{2 \over
{|\rv_i-\zeta|}}+V_{QW}(z),
\label{dz-hamil} \\
&V_{ee}(|\rv_1-\rv_2|)={2 \over |\rv_1-\rv_2|},
\label{vee}
\end{eqnarray}
where the vector potential is taken in the symmetric gauge
$\vec{A}={\rv \times \vec{B} /2}$.  The magnetic field is expressed in the
dimensionless quantity $\gamma={\hbar \omega_c /2R_y}$ with
$\omega_c={ eB/m^* c} $ the cyclotron frequency; $\zeta$ is the
position of the donor along the z-axis as measured from the center of
the well and $V_{QW}(z)$ is the confining potential due to the quantum
well of width W. For $GaAs/Al_x Ga_{1-x}As$ with $x=0.25$ we took
$\epsilon=12.5$ and obtain $a_B=98.7$\AA, $R_y=5.83\ meV$,
$\gamma=0.148 B(T)$. We took  the
mass of the 
electron equal in the well and in the barrier, namely
$m^*=0.067m_0$, and the
height of the barrier is given by $V_0=0.6 \times (1.155x+ 0.37x^2)\ eV$. 

The strong
confinement along the z-axis allows to neglect the correlation induced
by the Coulomb interaction
in the $z$-direction, thus we can write the wave functions for
the \dmn {} as follows
\begin{equation}
\Psi(\rv_1,\rv_2)=\psi(\vec{\rho_1},\vec{\rho_2})f_1(z_1)f_1(z_2),
\end{equation} 
with $f_1(z_i)$ the 1D ground state wave function for the electron
confined in a quantum well of height $V_0$.\cite{Messiah}

The two-electron function $\psi(\vec{\rho_1},\vec{\rho_2})$ expresses
the correlation between the two electrons and is obtained by
diagonalizing the Hamiltonian (\ref{dm-hamil}) in which  the
electron-electron, $V_{ee}$, and the electron-donor, $V_{ed}$, Coulomb
interaction are replaced by their average along the $z$-axis,
\begin{equation} V_{ee}(|\vec{\rho}_1-\vec{\rho}_2|)=\int{dz_1
\int{dz_2 |f_1(z_1)|^2|f_1(z_2)|^2 {2 \over
\sqrt{(\vec{\rho}_1-\vec{\rho}_2)^2-(z_1-z_2)^2}}}},
\label{e-e}
\end{equation}
and
\begin{equation}
V_{ed}(\vec{\rho})=\int{dz |f_1(z)|^2 {2 \over \sqrt{\rho^2 +
(z-\zeta)^2}}},
\label{e-d} 
\end{equation}
respectively. In a previous work \cite{platzman} it has been shown
that in the case of hard wall confinement Eq. (\ref{e-e}) can be
replaced by the expression
\begin{equation}
V_{ee}(|\rhv_1-\rhv_2|)={2\over {\sqrt{2\pi}} \lambda}
e^{|\rhv_1-\rhv_2|^2 / 4{\lambda}^2}K_0({|\rhv_1-\rhv_2|^2 \over
4{\lambda}^2}),
\label{approx} 
\end{equation}
 where $\lambda \cong 0.2W $  and $K_0 (x)$ is the modified Bessel
 function of the third kind. In the present paper we use the same
expression for a finite height quantum well in which
$\lambda$ is determined by fitting Eq. (\ref{approx}) to
Eq. (\ref{e-e}). A comparison between the potential 
(\ref{e-e}) that was evaluated numerically
and the approximate expression  
(\ref{approx}) is shown in Fig. \ref{pawfig}({\em a}) for a quantum
well of width $W= 200$\AA {} where the fitting parameter was found to be 
$\lambda=0.607a_B$.
  On the other hand, no simple analytic approximation to 
Eq. (\ref{e-d}) could be found. This is shown in
Fig. \ref{pawfig}({\em b}) for an off-center donor with $\zeta=0.7a_B$ and
$W=200$\AA {} where we compare Eq. (\ref{e-d}) which we fitted to 
the potential (\ref{approx})  with $\lambda=0.92a_B$ (solid curve) and the
screened Coulomb potential $ 1/ 
\sqrt{\rho^2+\lambda^2}$ with $\lambda=0.803a_B$ (dashed curve). None of
the two fits give a
good approximation to Eq. (\ref{e-d}) in the small $\rho$
region. Therefore, we retain in the Hamiltonian the numerical
expression for
Eq. (\ref{e-d}).
 
 Using a finite difference technique, as explained in Ref.
 12,  the Schr$\ddot{o}$dinger equation associated to the
 Hamiltonian  
 (\ref{dz-hamil}) was numerically solved on a non-uniform grid in
 $\vec{\rho}$-space and the eigenvalues and
 eigenvectors, $R_{n,l}(\rho)e^{i l \phi}$ 
 for the \dzr {} were
 found for different values of $\zeta$ and arbitrary magnetic field
 strength.  The eigenfunctions for the
 \dmn {} can then be constructed as a linear combination of the \dzr {}
 wave functions. Due to the rotational symmetry in the
 $\vec{\rho}$-plane of the Hamiltonian (\ref{dz-hamil}) the
 $z$-component of the orbital angular momentum, $L$,  is a good quantum
 number for those functions, and therefore the \dmn {} wave functions are taken
 as \begin{equation}
 \psi_L(\vec{\rho}_1,\vec{\rho}_2)=\sum_{k=1}^{k=k_m}\sum_{n=1}^{n
 =n_m}{\sum_{l=-l_m}^{l=l_m }}' C_{kn}^l
 R_{n,(L+l)/2}(\rho_1)R_{k,(L-l)/2}(\rho_2) e^{{i}[l(\phi_1-\phi_2)+L(\phi_1+\phi_2)]/2},
\label{wavefun}
\end{equation}
 where $\sum '$ indicates the summation is only over even (odd) values
 of the index $l$ when $L$ is even (odd).
 
\section{The Energy spectrum}
\label{enesp}
First we solve our model for an off-center donor in a
$GaAs/Al_{.3}Ga_{.7}As$ quantum well with width $W=200$\AA {} ($\approx
2a_B$) and height of the potential barrier $V_0=0.23eV$. The
dependence of the energy on the position of the 
 donor with respect to the center of the well is investigated numerically.

The binding energy of the \dmn {}   state with $\hat{z}$-component of the
orbital angular momentum equal to L is defined as

\begin{equation}
E_b^n(D^-,L)=E^0(D^0,0)+E(e,0)-E^n(D^-,L),
\label{binding}
\end{equation} 
 where $E^0(D^0,0)$ is the energy of the ground state of the \dzr {}
in the well, $E(e,0)=\gamma$ is the energy of a free electron in the $N=0$
Landau level  and $E^n(D^-,L)$ is the $n$-th energy
level of the \dmn 
{} with L the $\hat{z}$-component of the orbital angular momentum.

The results of our numerical calculation are plotted in
Fig. \ref{bind-energ} for $W=200$\AA. The binding energies of the first  
L=0 state, a
spin-singlet, and of the state L=-1, a spin-triplet, are plotted
against the magnetic field for different positions, $\zeta$,  of the donor with
respect to the center of the well.

We note, first, that the binding energy decreases when the donor center is  
displaced from the center of the quantum well. The reason is that
the electron-donor interaction decreases with
increasing $\zeta$. This is because,  due
to the strong confinement along the growth axis of the well, the electrons
tend, even in the case of an off-center donor system, to be
localized in the center of the quantum well although the donor is
displaced a distance $\zeta$ from the center.

A second feature to be noted is that the magnetic field dependence of the
binding energy changes qualitatively with increasing $\zeta$. For
sufficiently large $\zeta$ we find that  $E^n_b(D^-,L)$ has a maximum as
function of $\gamma$. The binding energy starts to decrease after this
maximum and for
sufficiently large $\gamma$  it can 
even become negative, indicating an unbinding of the \dmn {} state.
  
Third, in the absence of a magnetic field the ground state of the \dmn {} is,
regardless of the position of the donor, the spin-singlet state. When
increasing the magnetic field, the ground state for a well-center  \dmn {},
i.e. $\zeta=0$, remains the singlet one. In contrast, the
ground state 
of the off-center \dmn {} with $\zeta >.45a_B$ shows a transition to a
spin-triplet state for large enough  magnetic fields.  The magnetic
field at which the singlet-triplet 
transition occurs depends on the position of the donor as it appears
from Fig. \ref{bind-energ}. This
dependence will  be studied further below  where it is found that 
the magnetic field at which the transition occurs
decreases with increasing $\zeta$.
 
 Larsen and coworkers \cite{larsen} investigated the ideal 2D problem,
neglecting the finite extension of the electron wave function in the
z-direction, i.e. $f_1(z)=\delta(z)$, and calculated the \dmn {} spectrum
for a 
  donor out of the plane in the limit of high magnetic fields and
found an infinite number of 
  singlet-triplet transitions.
  The situation for a quasi-2D
  off-center \dmn {} is quite different. In this case, in contrast to the
2D case, the extension of the electron wave function in the z-direction
is taken into account, together with the finite height of the
barrier.  Let us  investigate deeper
  the behavior of the energy spectrum of such a system, with  
  e.g. $\zeta=0.7a_B$. The results for the binding energy of the different
  levels, i.e. different angular momentum states, are shown in
  Fig. \ref{.7}({\em a}), for  the case
  of a quantum well of width $W=200$\AA. Note that  different 
  transitions occur at higher magnetic fields. The ground state
  exhibits a singlet-triplet transition at $\gamma= 1.5$ and a
  triplet-singlet transition  at 
$\gamma=16.1$. 
For
$\gamma > 22.7$ which  corresponds to  $
B > 154\, T  $  the \dmn {} ground state unbinds,  
i.e. the
  \dmn {} {\it magnetically evaporates}.

While for the ideal 2D system an infinite number
of singlet-triplet transitions are found for a quasi-2D system only a
finite number of such transitions are possible as is clearly visible
from Fig. \ref{.7}({\em a,b}). The critical $\gamma$'s
at which the singlet $\leftrightarrow$ triplet transitions occur depend on the
position of the donor (see Fig. \ref{.7}({\em b})). The $\gamma -
\zeta$ phase diagram for the
ground state of a quantum well of width $W=200$\AA {}  is given in
Fig. \ref{g.vs.zeta}. We found that for $\zeta < 0.45a_B $  the ground
state is a singlet for all magnetic fields, for  $0.45a_B <\zeta < 0.65a_B$ 
 only one singlet-triplet transition (see
Fig. \ref{g.vs.zeta}) is possible and for $ \zeta > 0.65a_B$ there are two
of such  transitions.  Increasing $\zeta$ further 
such that the donor is in the 
  barrier (i.e. $\zeta > 1.01a_B$), the number of singlet-triplet
  transition does not  increase
as illustrated in  Fig. \ref{.7}({\em b}) for a \dmn {} with  $ \zeta=1.4a_B$ .
 
 The physical origin of the  singlet-triplet transitions  is related
 to the decrease of the 
electron-donor attraction with the displacement of the donor from the
center of the well when  compared to the constant electron-electron
repulsion. The corresponding electron-donor
and electron-electron in-plane potentials are shown in
Fig. \ref{repul-attrac} for two values of $ \zeta $. For small values of
$\zeta$ ( e.g. $ \zeta =0. $ in  Fig. \ref{repul-attrac}) the attractive
single electron-donor potential is larger than the electron-electron
potential and consequently the \dmn {} system  prefers a
configuration  in which  the 
two electrons are as close as possible to the donor in order to enhance
the binding energy, i.e. the L=0 state is favored. When $ \zeta $ is
sufficiently large (e.g. $ \zeta=0.7a_B$ in  Fig. \ref{repul-attrac}) the
repulsive electron-electron interaction dominates  the attractive
single-donor potential at small distances and the \dmn {} can have
bound states only when the two electrons are sufficiently apart
to render   the repulsive inter-electrons interaction lower or of the
same order as the attractive electron-donor potential. For small
magnetic fields  this can still be realized in the L=0
state. Increasing the magnetic field  brings the electrons closer to
$\rho=0$ which will also increase the electron-electron repulsive
energy. For sufficiently small $\zeta$ this can still be compensated by
the attractive electron-donor energy. For $\zeta$ sufficiently
large the electron-electron repulsive energy increases faster
then the electron-donor energy with increasing $B$. The \dmn {} system
can decrease its energy in this case by placing the electrons
further apart which is
achieved by placing the electrons in higher L-states. Similar
singlet-triplet transitions have been  found recently in quantum dots
systems.\cite{magic,Twodimqd,ultima} The quantum dot system is an
extreme case in which the electron-donor potential is replaced by the
confinement potential which is  usually taken of a quadratic form,
i.e. $V_{ed} \rightarrow \omega^2 \rho^2$.

In Fig. \ref{correlation} the 
pair-correlation function $  <\delta(\rho -
|\vec{\rho}_1-\vec{\rho}_2|)> $  is shown for the spin-singlet L=0 (
Fig. \ref{correlation}({\em a})) and
spin-triplet L=-1 ( Fig. \ref{correlation}({\em b})) states for
different values of the magnetic field for
an on-center 
(i.e. $ \zeta=0$)  and for an
off-center (i.e. $ \zeta=0.7a_B$) 
\dmn {} system. The magnetic field behavior  of the
two states is essentially the same for both the center and 
the off-center \dmn {} system. The magnetic field tends to localize more the
wave function with increasing magnetic field. For the L=0 state the pair
correlation function becomes 
more and more peaked at $\rho=0$, this means that the electrons are more
and more close to each other with increasing magnetic fields. For the
L=-1, the peak of the 
correlation function is shifted towards $ \rho=0 $ with increasing $B$,
and thus, the 
magnetic field localizes the electrons further. The effect of the
electron-electron repulsion can be seen in the shape of the correlation
function itself. For the off-center system the pair-correlation
function is broader than the one for the center \dmn {} system  even
for increasing 
  magnetic fields and  thus the electrons tend to repel each other
  more, which is a consequence of the diminished electron-donor interaction.       

When the dimension of the well is reduced the localization of the
electrons in the center of the well is increased. For example, 
 if we neglect the penetration of the electrons in
the barrier, the width of the $ f_1(z)$ is equal to $ L$.  Thus the
electron-electron repulsion increases and at the same time, for the
off-center case, the 
electron spends more time far from the position where the donor is
located. Thus,  we expect that systems
with a smaller well width  will show more spin-singlet to 
spin-triplet transitions with increasing magnetic field, and that
these transitions will  occur at
smaller fields.
 
 Indeed, for a $ W=100$\AA {} quantum well, with again $\zeta=0.7a_B$, we
 observe (see Fig. \ref{100.7}) as much as 4 transitions before the
\dmn {} evaporates at a 
 magnetic field of  $ B \approx 81 T $ (i.e. $ \gamma \approx 12.0 $). The full
phase-diagram for those
transitions is shown in Fig. \ref{phasediag II}.  The
well  width dependence  of the singlet-triplet transitions and of the
evaporation magnetic field are shown in 
 Fig. \ref{wellw} for $\zeta= 0.7 a_B$. Notice that the critical magnetic field for the same
 transitions, e.g. for the spin-singlet L=0 to spin-triplet L=-1 state,
 decreases with decreasing well width. At the same time the number of
 transitions increases. But the evaporation magnetic field first
 decreases and then for $W < 140$\AA {} increases again. An explanation of this feature is
 that other transitions are allowed for small well width
 and this ensure stability of the \dmn {} up to higher magnetic fields.

 The increase of the number of singlet-triplet transitions with 
 decreasing  dimensions of the quantum well, explains the larger number of
 transitions found by Larsen {\em et al.} \cite{larsen} in the ideal 2D
 system with 
respect to the smaller
 number found in the present  study of realistic quasi-2D systems.

\section{Cyclotron resonance transitions}
\label{fo}


The oscillator strength for cyclotron transitions, in the present units, is
defined as
\begin{equation}
F_{i,f}=(E_f-E_i) | < \Psi_i | \sum^2_{j=1} {1 \over 2} \rho_j e^{\pm
i \phi_j} | \Psi _f > |^2
\label{oscillator}
\end{equation}
where $E_f$, $E_i$ are, respectively, the final and initial-state
energies and  $\psi_f $,
$\psi_i$ are, respectively, the final and initial-state wave
functions. The $\pm$ sign in Eq. (\ref{oscillator}) refers to circular
left/right  polarization of the light.  Note that the perturbation induced by the electric field
is spin-independent,
thus the initial and final-states conserve the total spin, i.e. they are both
spin-triplet or both spin-singlet states. Eq. (\ref{oscillator}) leads
to the
selection rules $ \Delta L = \pm 1 $, 
while no
selection rule is present for the quantum number n.

We have studied the oscillator strength for cyclotron 
resonance transitions from
the first singlet L=0 state - (n,L,S)=(1,0,0) -  to the (1,-1,0) and 
the (1,1,0) states in the range 2-15T. The transition energies and
oscillator strengths, 
for $\zeta=0.7a_B$, 
 are plotted in Fig. \ref{tran-sin} against the magnetic field and are
 compared to the one for $\zeta=0.$ We
recall that for a two-electron atom the oscillator strength satisfies
the sum-rule $ \sum_i F_{i,f}=2 $. We observe that the off-center
and the  center \dmn {} have  rather similar
qualitative
magnetic field dependences.


 Cyclotron
resonance transition from the ground state should show a
discontinuous behavior in the cyclotron 
transition energies as a function of the
magnetic field at the singlet-triplet transition points. In Fig. \ref{85} the transition energies for a donor
at 
position $ \zeta=0.85a_B$ are shown. Respectively, the transition
energies for 
$(1,0,0) \rightarrow (1,1,0)$ and $(1,-1,1) \rightarrow (1,0,1)$ and 
$(1,-2,0) \rightarrow (1,-1,0)$ are plotted. The solid curve shows 
the transition energy which we expect to observe if the system makes
a cyclotron resonance transition starting from the ground state. Thus 
 steps in the cyclotron resonance energy should be
observed at those magnetic fields
at which the singlet-triplet transition takes
place. In  real experiments, as
we will see later, not always transitions only from the ground state are
seen in the neighborhood of the critical field which is due to the fact that
in a real experiment the temperature is non zero. Indeed, when the
 {\em old}  ground-state, i.e. the state that was
before the transition the ground-state,  and the {\em new}
ground-state, i.e. the state that is after the transition the
ground-state,  have a 
comparable binding energy they can both be thermally populated.   
   
\section{Comparison with experiment}
\label{espdata}

In this section we present a comparison between our theoretical results and the
experimental data reported by Jiang {\it et al.}\cite{McCombe}
The experiment of Jiang {\em et al.} was performed on multi-layers of
$GaAs/Al_{.3} Ga_{.7}As$ with well width of 200\AA {} and 
barrier width of 600\AA. Such a system can be considered 
as an ensemble of single quantum wells. 
The wells were nominally $ \delta$-doped at 3/4 of the distance between
the center of the well and its edge. In the model discussed in this
paper this means that $ \zeta=.75a_B$ .
 
 A comparison between the theoretical and the experimental transition
energies is reported in Fig. \ref{expdata}. The observed transitions
are the \dzr {} $ (1,0)\rightarrow (1,1)$ and the \dmn {} singlet $(1,0,0)
 \rightarrow 
(1,1,0) $ and triplet $(1,-1,1) \rightarrow (1,0,1)$ transitions. Our
theoretical results are given by the three different curves. Note
that our results fit well the
data at low magnetic fields. The deviations between theory and
experiment observed for $B > 9 T$ can be attributed to band
non-parabolicity and polaron effects. Both effects decrease the
transition energy\cite{Shi} but are not taken in account in
this paper.

 
 In cyclotron resonance experiments the integrated absorption intensities can be
measured.  The integrated absorption intensities are proportional to
the oscillator strength times the population
densities of the levels involved in the transition.
To compare our results with the experimental data we have to make an
assumption on the form of the population density.
 We assume that only the initial level of the transition is
populated. Thus for the off-center \dmn {} the population density of
the level is proportional to $ e^{E_b/ kT}$, where $E_b$ is the binding
energy of the initial state.  We remark that for off center \dmn {}
in this range of magnetic fields the energies of the triplet and the
singlet states are comparable.
For the well-center \dmn, instead, we consider only the L=0
spin-singlet state to be populated, i.e. the population density is 1.

The results for the relative integrated intensities of the singlet transition 
as evaluated in our
calculation and the experimental results are plotted in
Fig. \ref{oscill} and are in good agreement,
both for the well-center as well as  for the off-center \dmn {}.
The temperature in the experiment was $T=4.3K$. Note
that the different magnetic field dependence for the center \dmn {}
(i.e. increase with B) and for the off-center \dmn {}
(i.e. decrease with B) is correctly described. The errors bars for
the off-center intensities are rather large. A slight
discrepancy is observed at certain values of the 
magnetic field for the off-center \dmn {} but we observe that moving
the donor in our model slightly closer to the center of the well,
i.e. $ \zeta =0.7a_B$,
   the relative integrated intensity changes from the solid to the dotted curve
in Fig. \ref{oscill} and now matches the experimental data in the
magnetic fields 
region in which there was not such a  good agreement before. Thus 
the apparent discrepancy  in the integrated
intensity, with $\zeta=0.75a_B$, in the range $5-14\ T$
is explained by considering a small distribution
of donors around the point of intended $\delta$-doping.

\section{Summary and Conclusion}
\label{conclusion}

We presented a theoretical study of the off-center \dmn, where
special attention was paid to the dependence
of the binding energy on  the well width and the donor position. 
We found that the magnetic field
induces spin-singlet to spin-triplet transitions in the ground-state
of the off-center \dmn. The number of those transitions depends
{\em both}
on the position of the donor and on the width of the well.
In contrast to  the ideal 2D system and to quantum dots only a
finite number of transitions are found. If the donor is near the
center of the quantum well  no such singlet-triplet transitions occur.
When such singlet-triplet transitions occur we find that at
sufficiently large magnetic field the \dmn {} system becomes unbound
and consequently one observes a magnetic evaporation of the \dmn {} system.
We calculated also the oscillator strength for the off-center \dmn {} as
function of  the magnetic field and compared it to the results for a center
\dmn. We restrain ourselves to the study of the optical transitions
$(1,0,0) \rightarrow (1,-1,0)$, $(1,0,0) \rightarrow (1,1,0)$,
and we observed that the off-center and the center \dmn {} have
similar magnetic behaviour.
Our results were used to explain the experimental results
recently reported by Jiang {\it et al.} on the cyclotron resonance
transition energy and the absorption intensity of the off-center \dmn {}
system for magnetic fields up to $ 15 T$. 

In conclusion, the \dmn {} center is a natural quantum dot system
which is confined by the coulomb potential of the impurity and
consequently is more closely related to real atomic systems. A
remarkable feature of the \dmn {} centers in quantum wells is the
controllability of the effective confinement potential which is
Platzmann-like for a donor in the center of the well and screened
Coulomb-like when the donor is placed far away from the quantum well
center.
In the latter case the potential is parabolic near the center of the
quantum well plane and thus resembles the confinement potential of
quantum dots. In this case singlet-triplet transitions are found as
function of the magnetic field. A crucial difference with the quantum
dots is that only a finite number of such a transitions occur and that
for sufficiently large magnetic fields the \dmn {} system becomes
unbound, i.e. magnetically evaporates.

\section{Acknowledgment}

Part of this work is supported by the EC-programme INTAS-93-1495-ext, the
Russian Foundation for Basic research 95-02-04704, the Flemish Science
Foundation ( FWO-Vl) and the `Interuniversity Poles of Attraction
Program - Belgian State, Prime Minister's Office - Federal Office for
Scientific, Technical and cultural Affairs'. F.M.P. is a Research
Director with FWO-Vl.

\begin{figure}
\caption{Comparison between the numerical evaluation and the
  analytical fitting of the average in-plane potentials for a
  quantum well of width $ W=200$\AA.  In ({\em a}) the e-e
  potential (\ref{e-e}) is fitted to Eq. (\ref{approx})
  (dashed curve).  In ({\em b}) the e-d potential (\ref{e-d}) is
  fitted to Eq. (\ref{approx}) (solid curve) and to
  $ 1/ \sqrt{\rho^2+\lambda^2} $ (dashed curve).} 
\label{pawfig}
\end{figure}
\begin{figure}
\caption{The magnetic field dependence of the L=0 spin-singlet binding
  energy (solid curves)  and the L=-1
spin-triplet binding energy (dotted curves) for a $  GaAs/Al_{.3}Ga_{.7}As $ quantum
well with width $ W= 200 \mbox{\AA}=2.02a_{B} $ are shown for different
position, $\zeta$,
of the donor with respect to the center of the well. $ \zeta $ is in
units of $a_B$.  For increasing magnetic field there is
a crossing between the spin-singlet and the spin-triplet states when $ \zeta
 > 0.45 a_B $. .  }
\label{bind-energ}
\end{figure}

\begin{figure}
\caption{In ({\em a}) the binding energies for different values of the
z-component of the angular momentum are shown for a \dmn {} with the
donor placed at  $\zeta=.7 a_b \approx 70$\AA {} from the center of the
quantum well. In ({\em b}) the binding
energies for a barrier \dmn {} are displayed, with $\zeta=1.4 a_B
\approx 140$\AA.}
\label{.7}
\end{figure}

\begin{figure}
\caption{Phase diagram for a quantum well of width $W=200$
  \AA. The curves  show the magnetic fields at which  the 
  singlet-triplet transitions occur for given position of the donor as
  well as the field at which the \dmn {} system evaporates.
 }
\label{g.vs.zeta}
\end{figure}

\begin{figure}
\caption{ The in-plane electron-donor potential for the well-center
\dmn {} and for the off-center \dmn {} are compared to the
electron-electron potential.}
\label{repul-attrac}
\end{figure}
\begin{figure}

\caption{ The pair correlation function of the \dmn.
In ({\em a}) the correlation function of the spin-singlet L=0 state is
presented for different values of the magnetic field both for the
off-center (dotted curves) as well as  for the center \dmn {} (solid
curves). In ({\em b}) the same plot as in 
({\em a}) is made but now for the spin-triplet L=-1 state.}
\label{correlation}  
\end{figure}

\begin{figure}
\caption{ The binding energies for a \dmn {} with $\zeta=0.7a_B$ in a
100\AA {} quantum well, for different values of L.  Four
transitions occur 
before the \dmn {} system evaporates which occurs for $\gamma=13.5$.}
\label{100.7}
\end{figure}

\begin{figure}
\caption{ The phase diagram for  a 100\AA {} wide quantum well.
 Four transitions are possible for this quantum well.}
\label{phasediag II}
\end{figure}

\begin{figure}
\caption{ The phase diagram for fixed donor position, $\zeta=0.7a_B$,
  as function of the well-width $W$}
\label{wellw}
\end{figure}

\begin{figure}
\caption{The transition energies, ({\em a}), and the oscillator
  strengths, ({\em b}), for  the $
(1,0,0)\rightarrow (1,-1,0)$ and for the $(1,0,0) \rightarrow (1,1,0)$
transitions. The values for the donor placed at $\zeta=0.7a_B$ are compared to the values
for a well-center \dmn {}. The dotted line is the free electron
cyclotron transition energy, $ \hbar \omega_c$.}
\label{tran-sin}  
\end{figure}


\begin{figure}
\caption{ The cyclotron resonance transition energies for a donor at $\zeta=0.85a_B$ are
  shown by the dashed  curves for the first three lowest
  states. The solid curve represents the expected transition energy
  from the ground state as
  function of the magnetic field at zero temperature.}
\label{85}
\end{figure} 

\begin{figure}
\caption{The experimental data of Jiang {\it et al.} {\cite{McCombe}}
  for the cyclotron resonance transition energy (symbols) are compared
  to our theoretical results (curves), for the \dzr {} and the singlet
  and triplet \dmn {}. The donor is at $\zeta=0.75a_B$ and the well
  width is $W=200$\AA.}
\label{expdata}
\end{figure}

\begin{figure}
\caption{Comparison of the relative integrate absorption intensity
between the experimentally measured (symbols) and the present theoretical
results (curves). The donor position is at 
$\zeta=0.75a_B$.  The dotted curve takes into account a displacement of the donor from
the position at which the well is nominally $ \delta $-doped,
i.e. $\zeta=0.7a_B$.} 
\label{oscill}
\end{figure} 

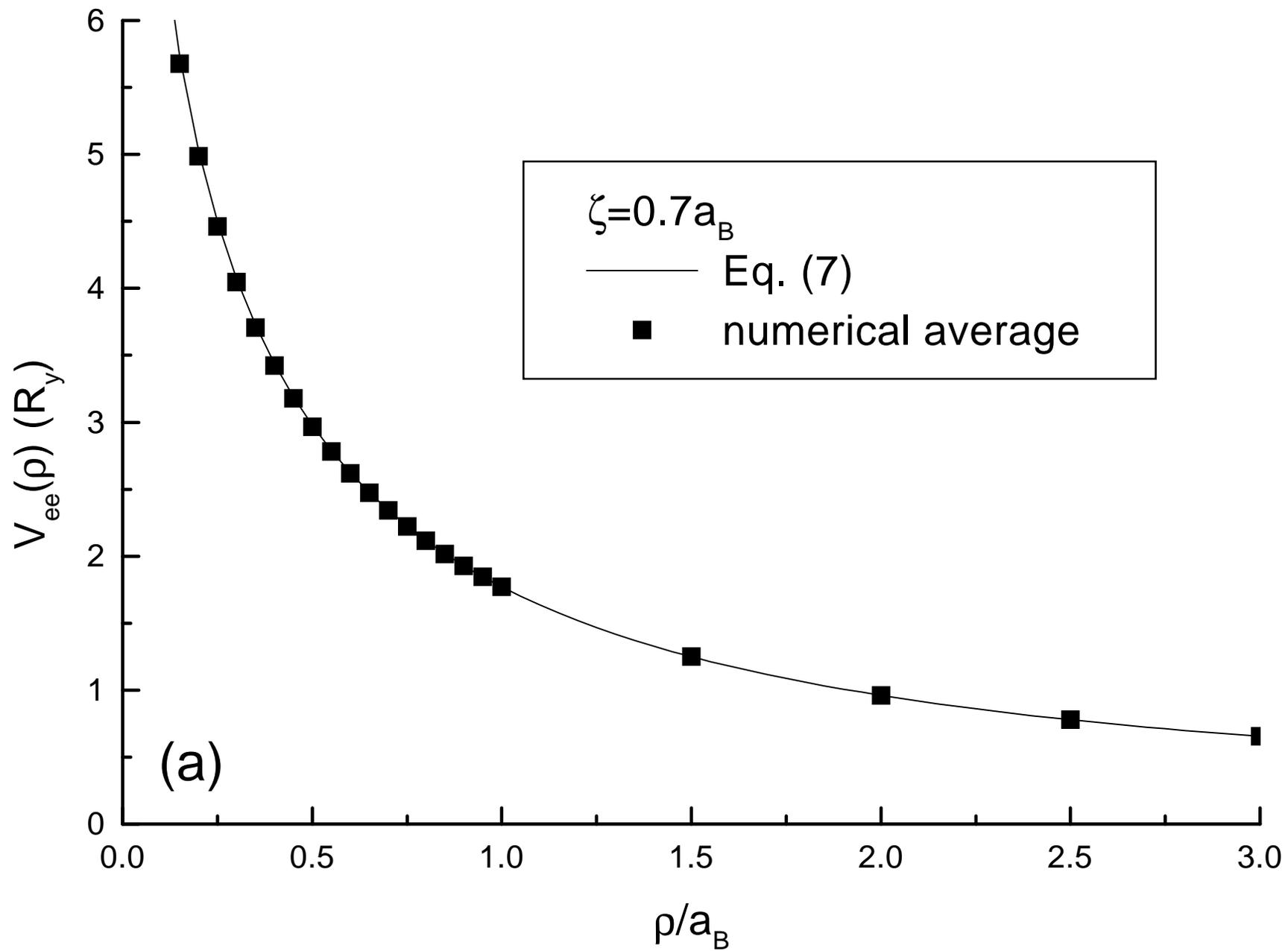
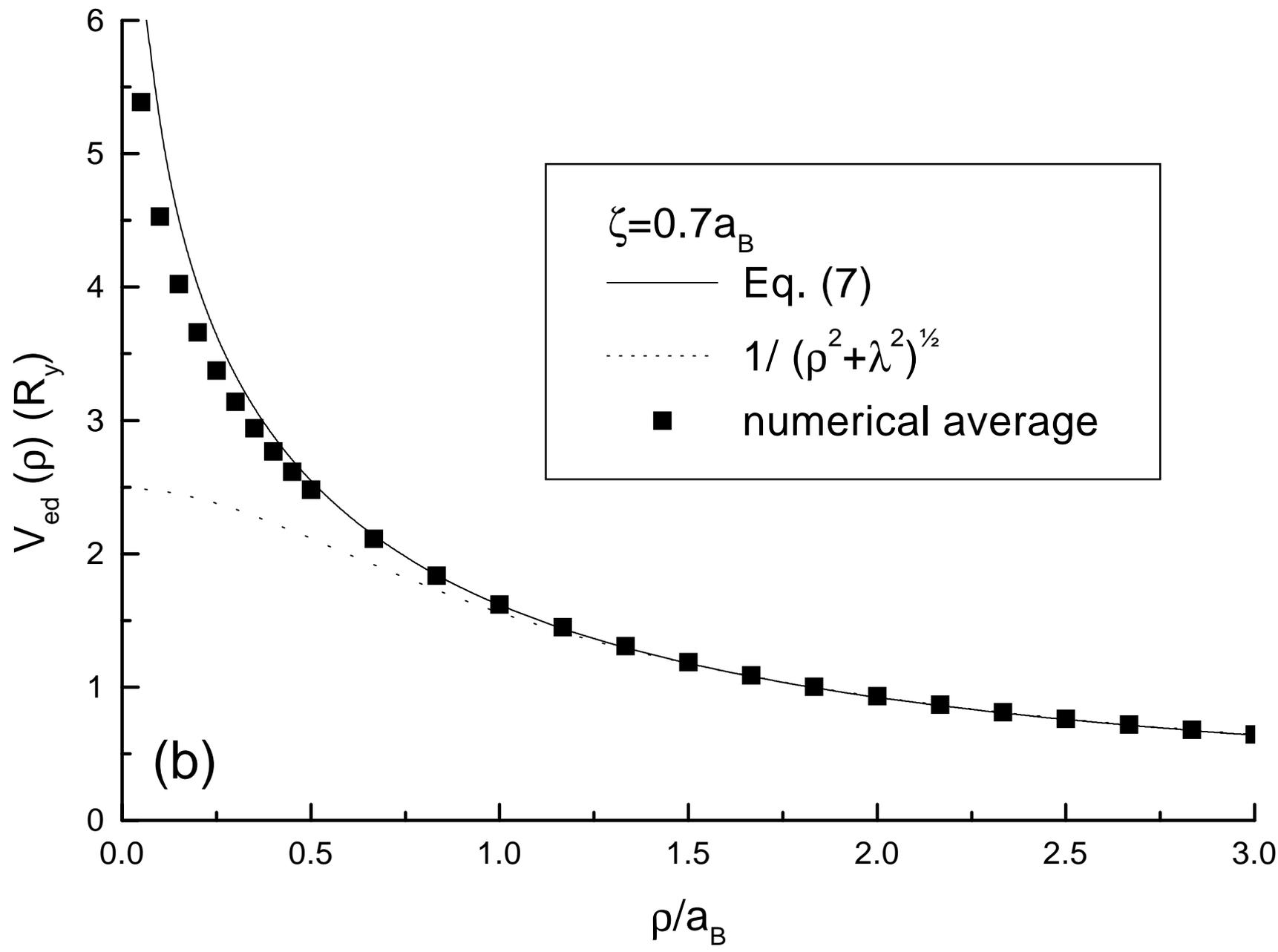
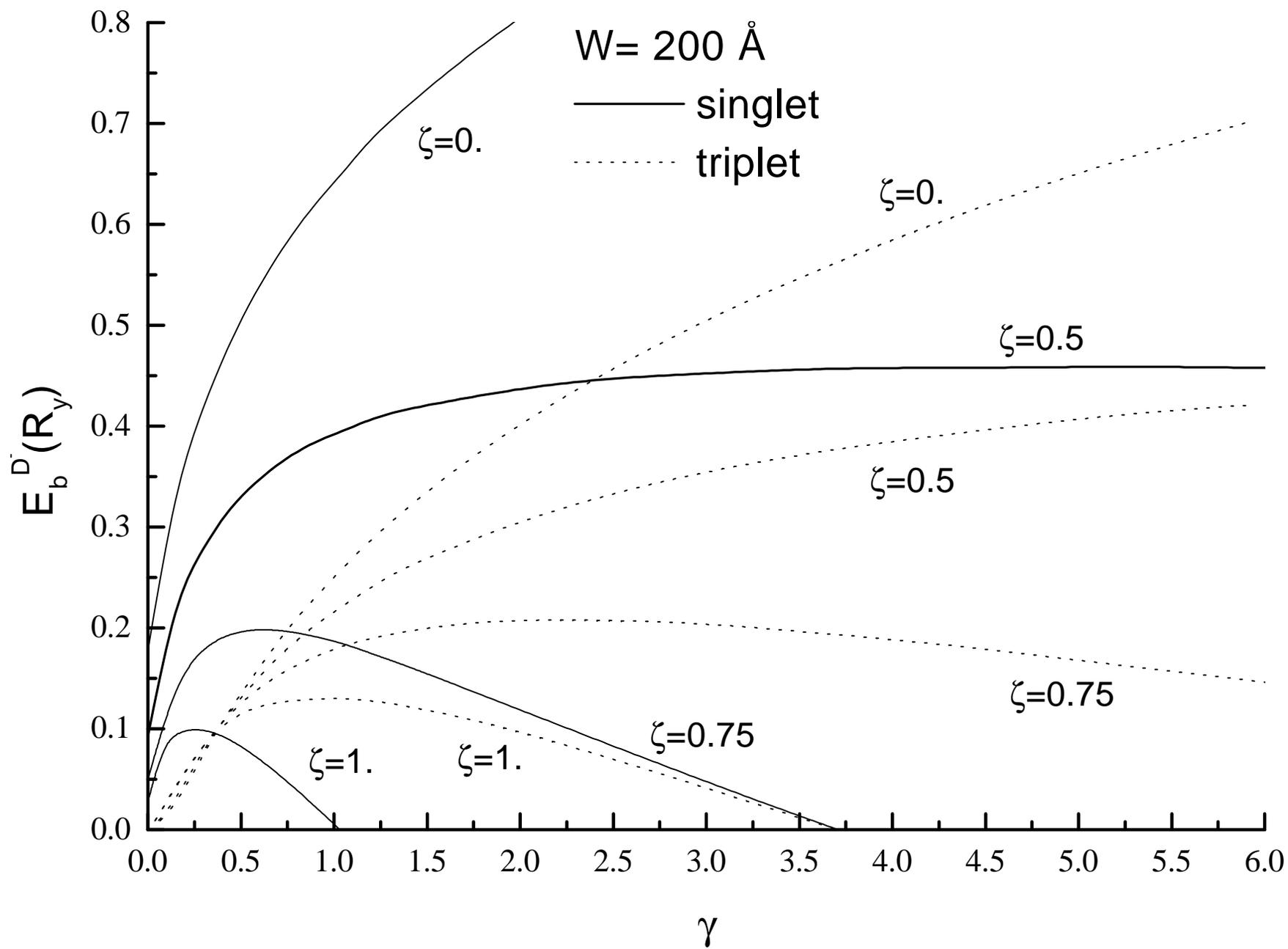
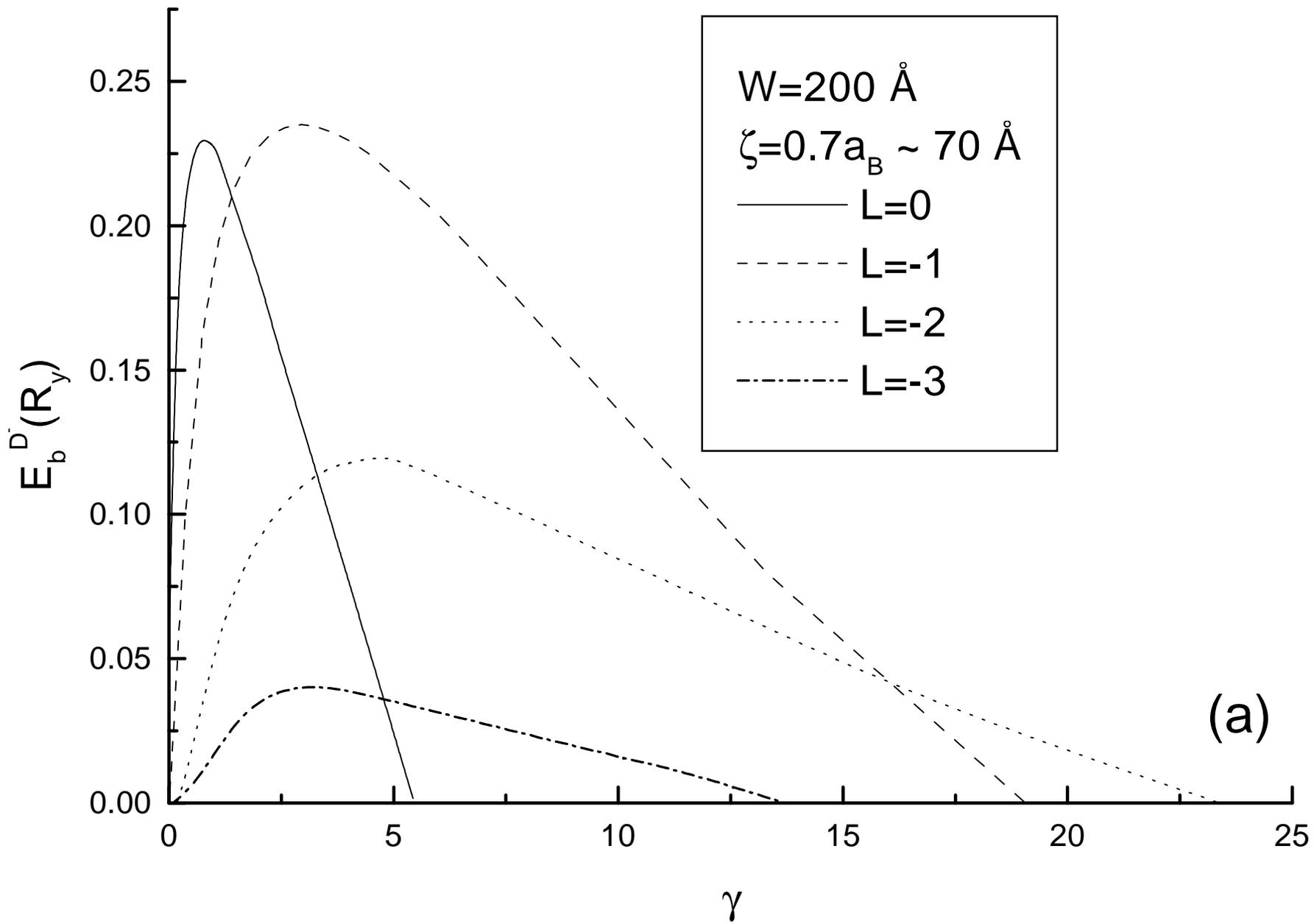
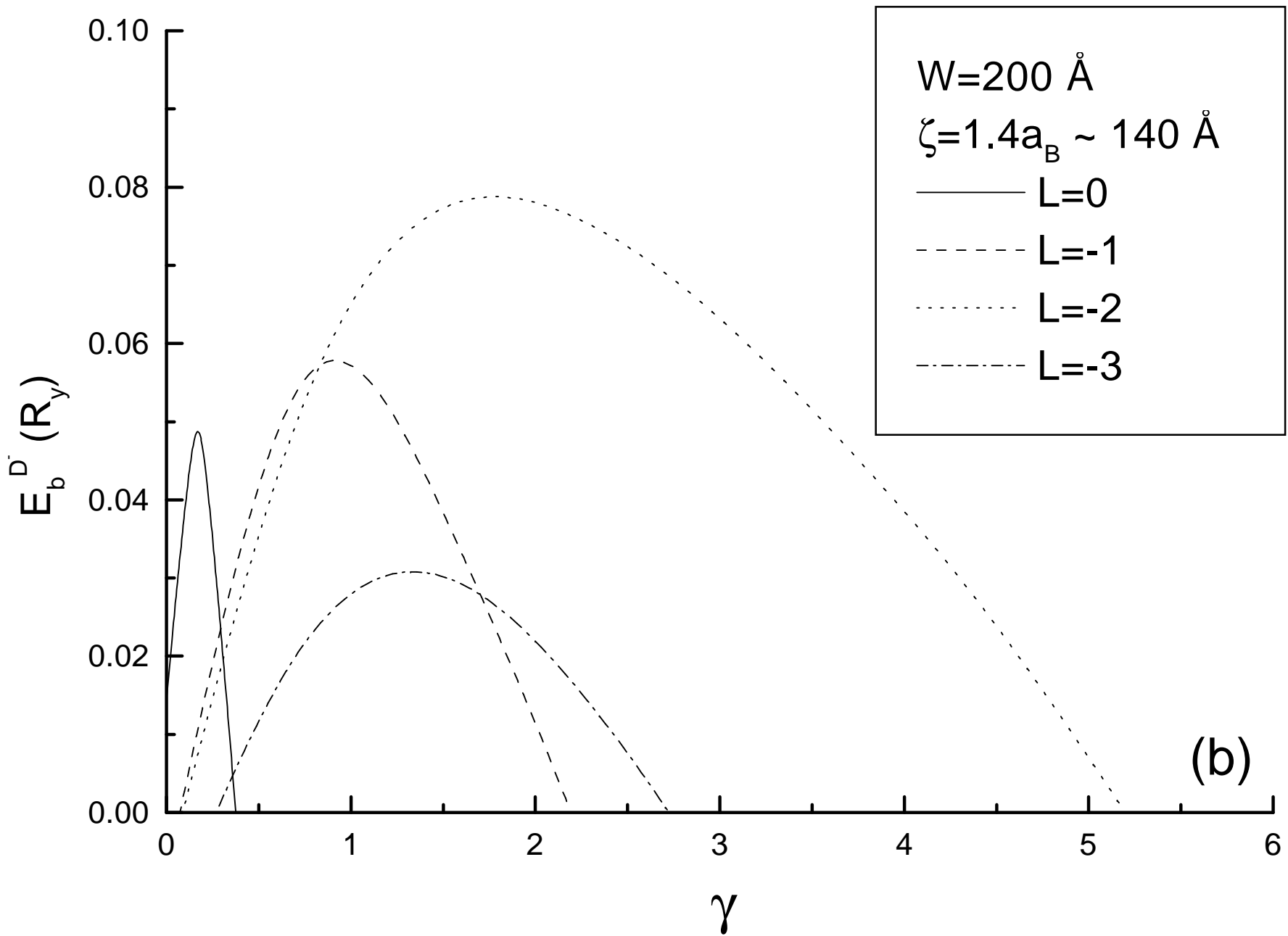
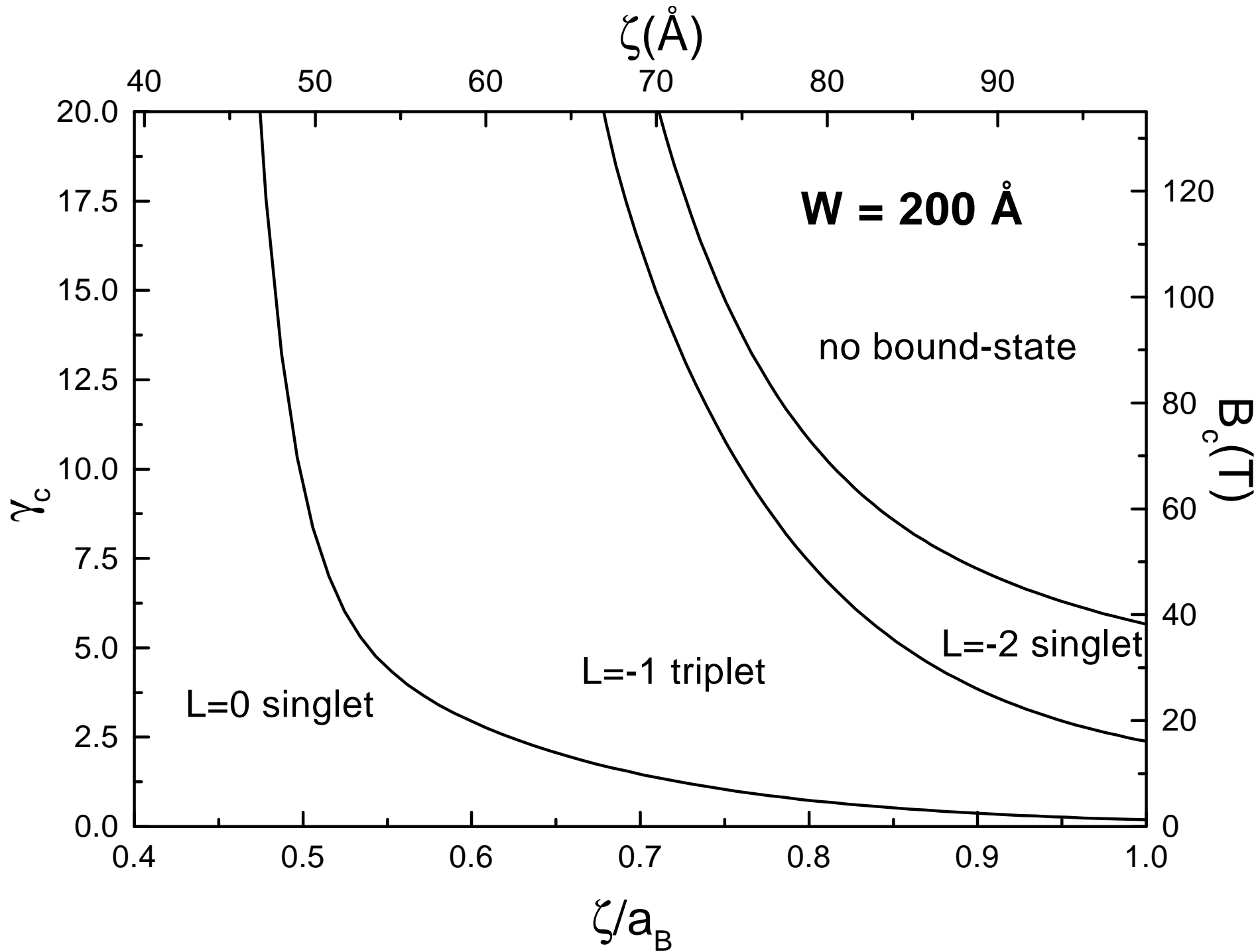
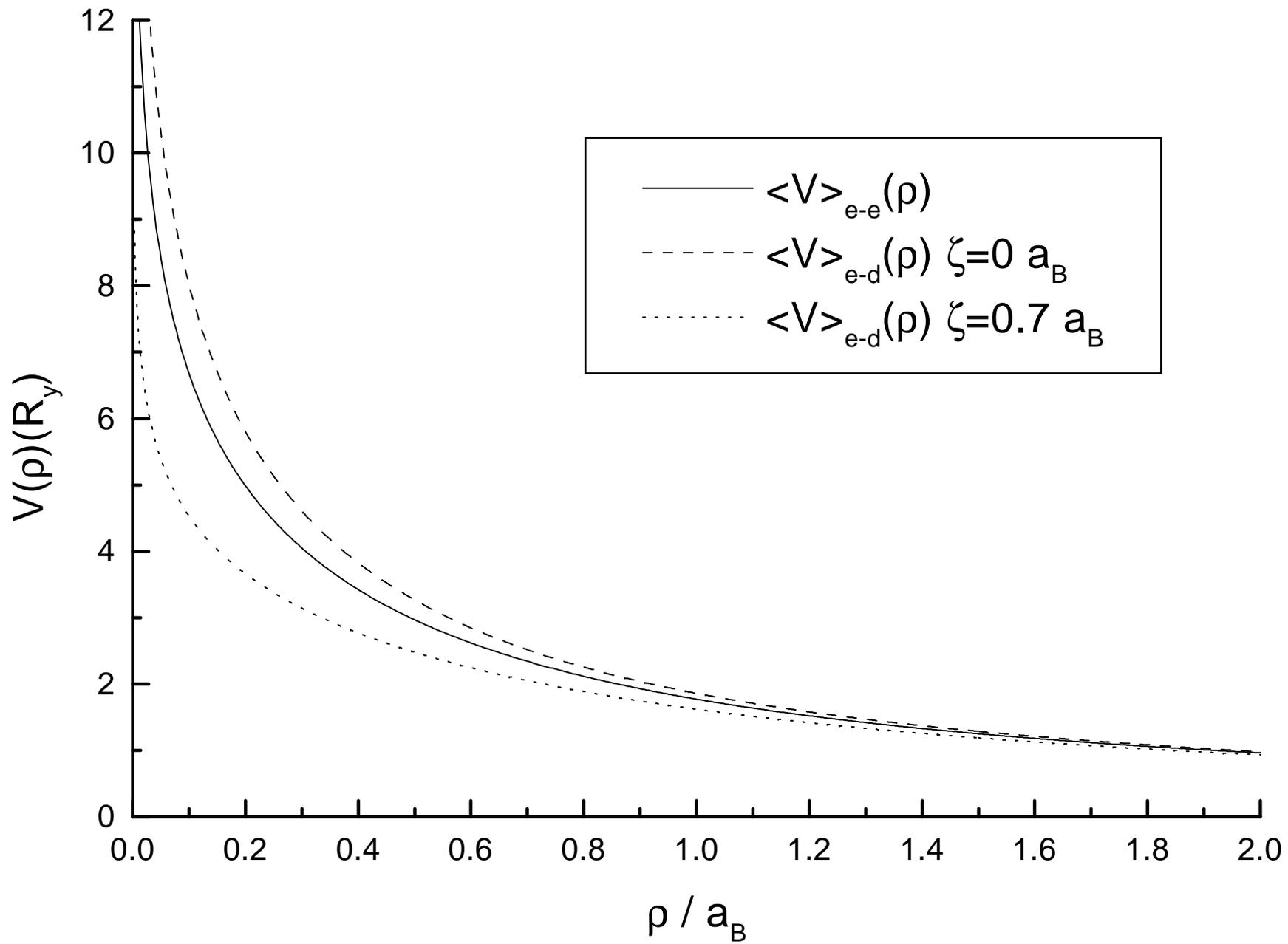
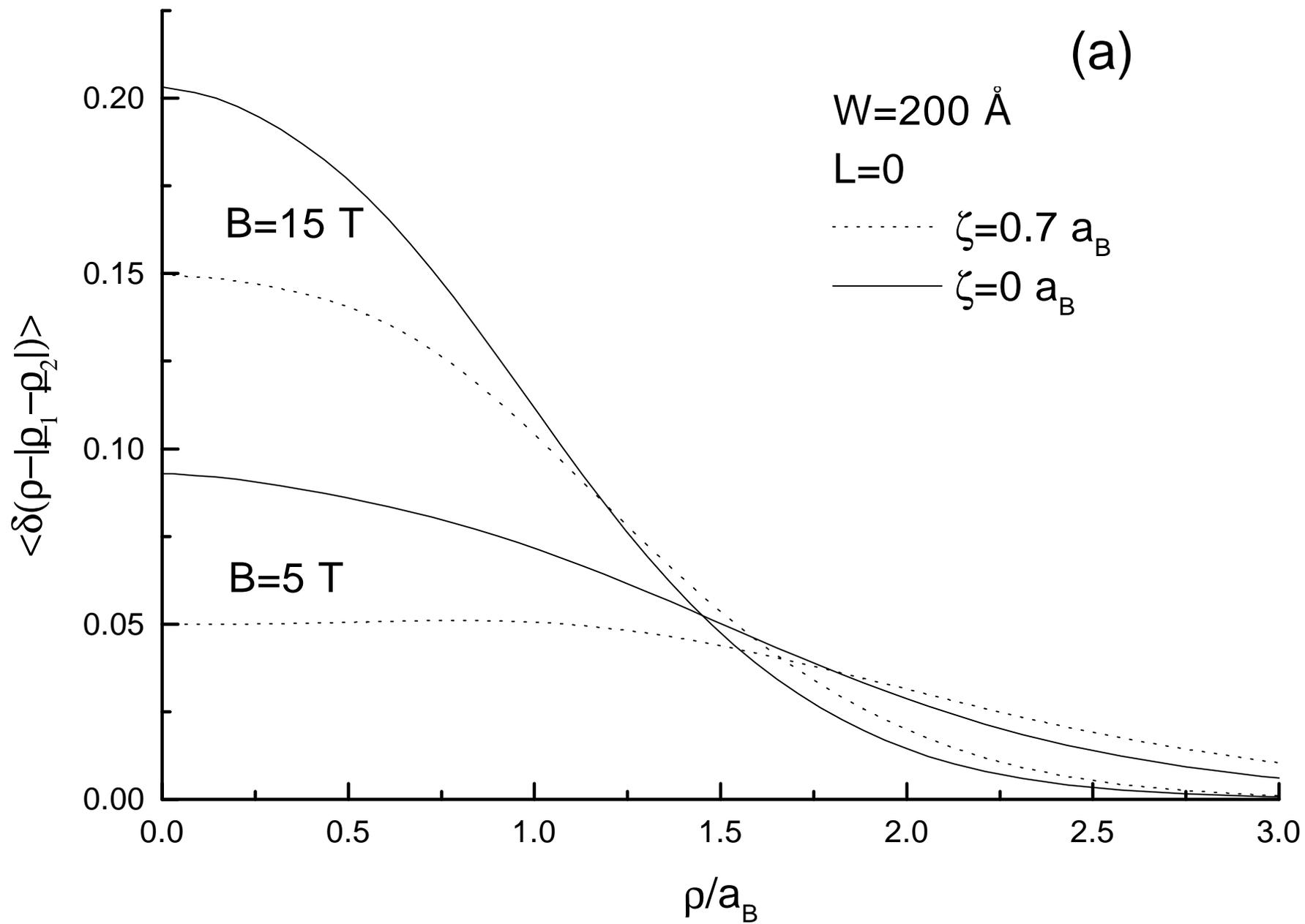
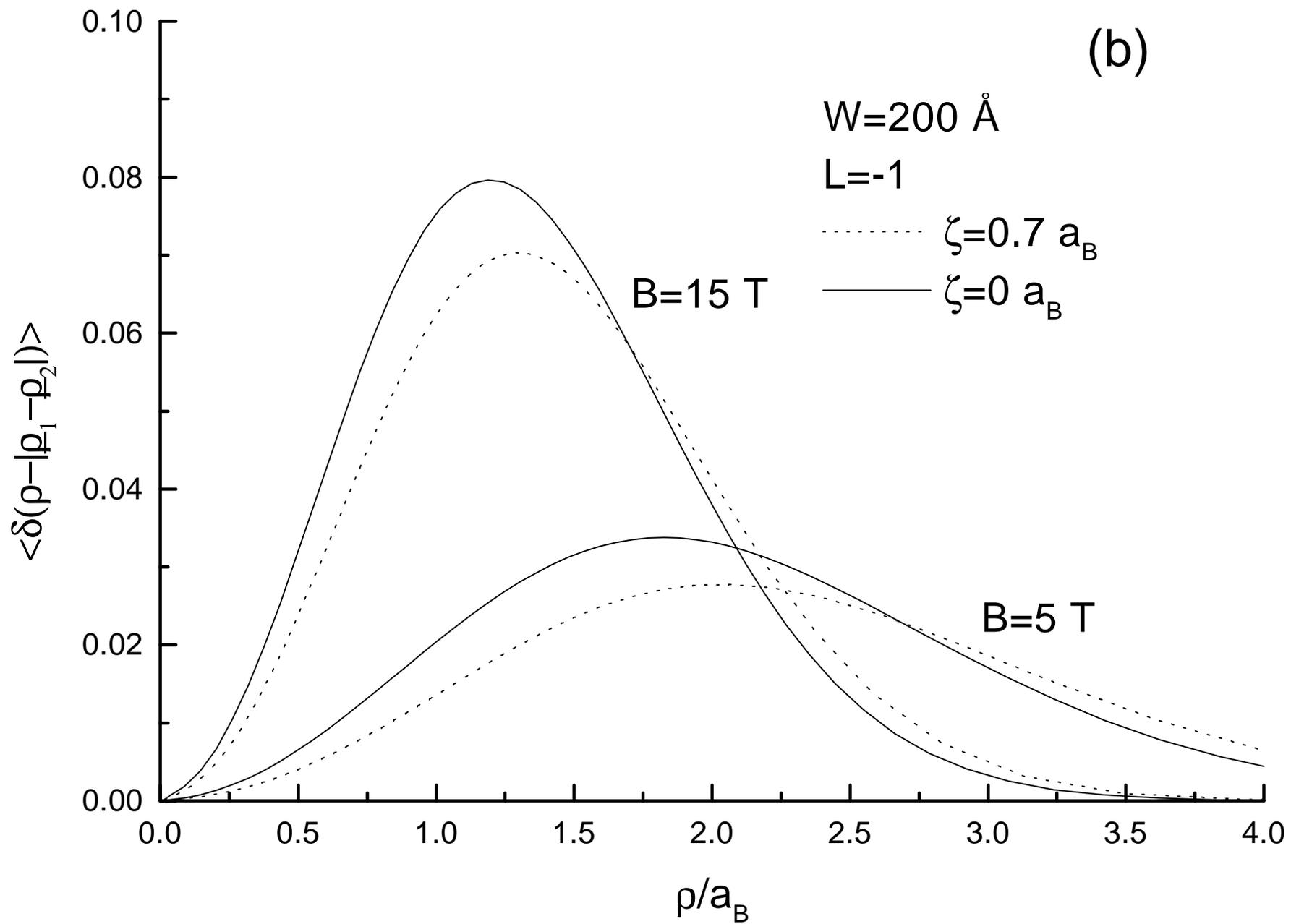
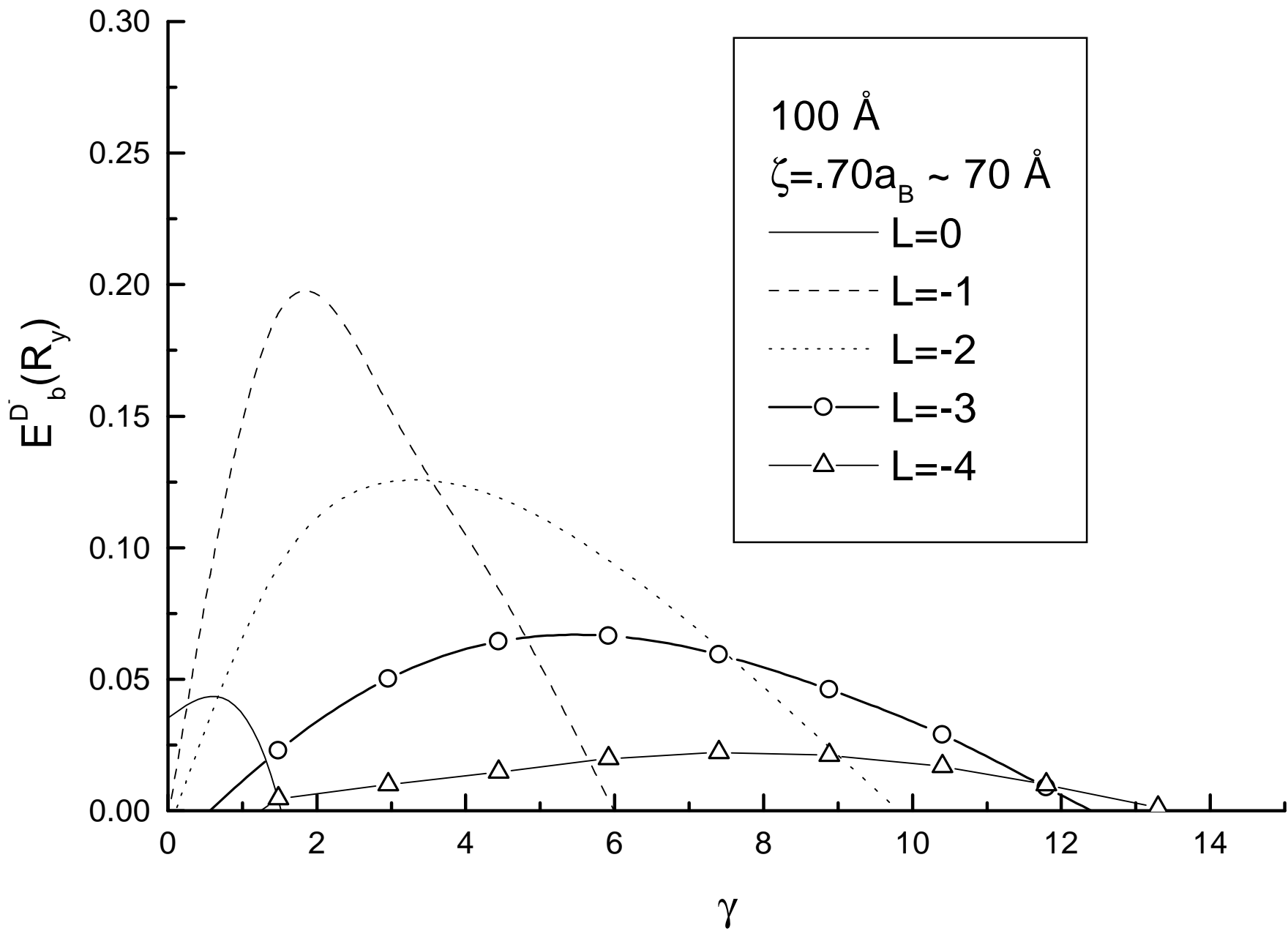
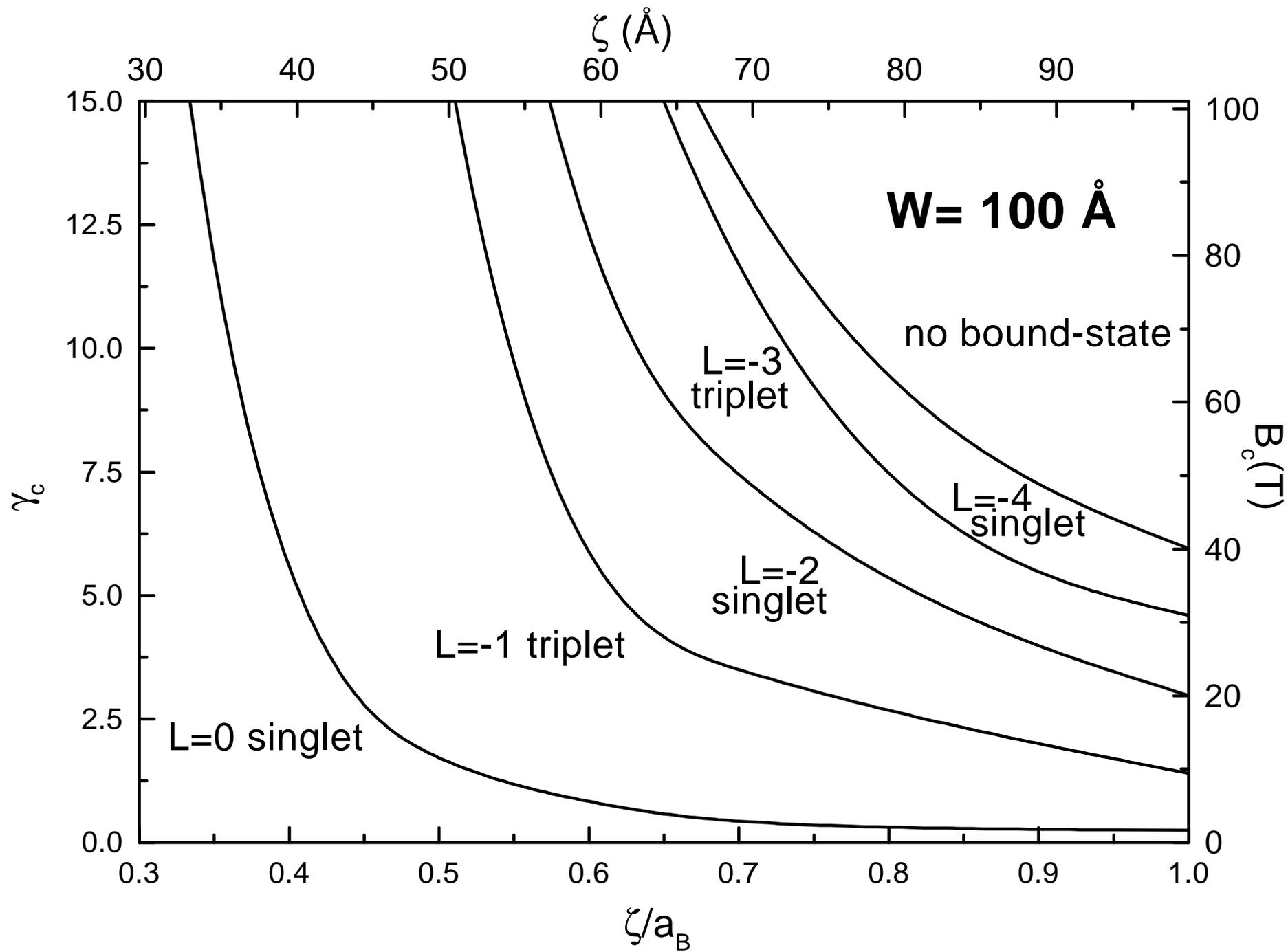
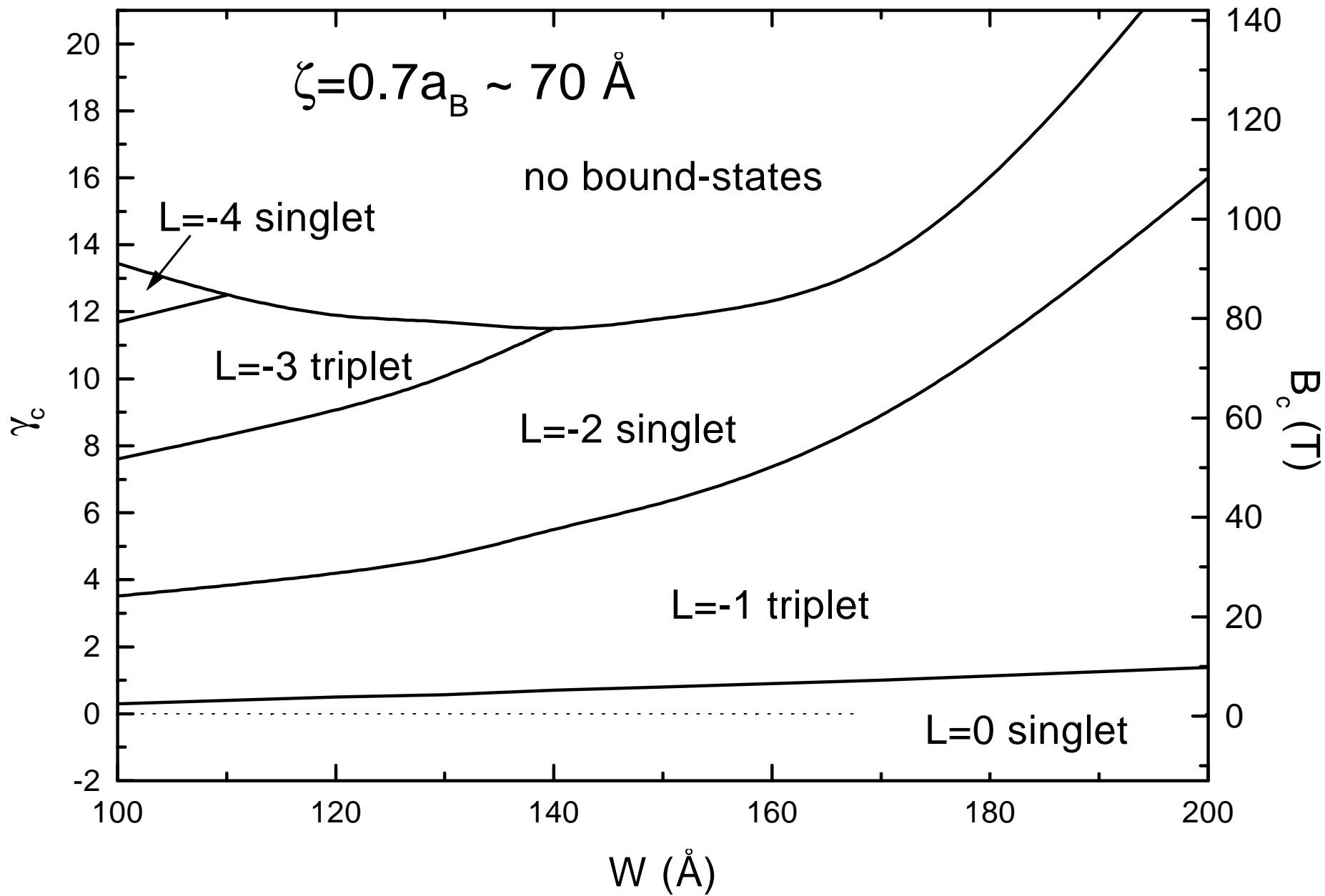
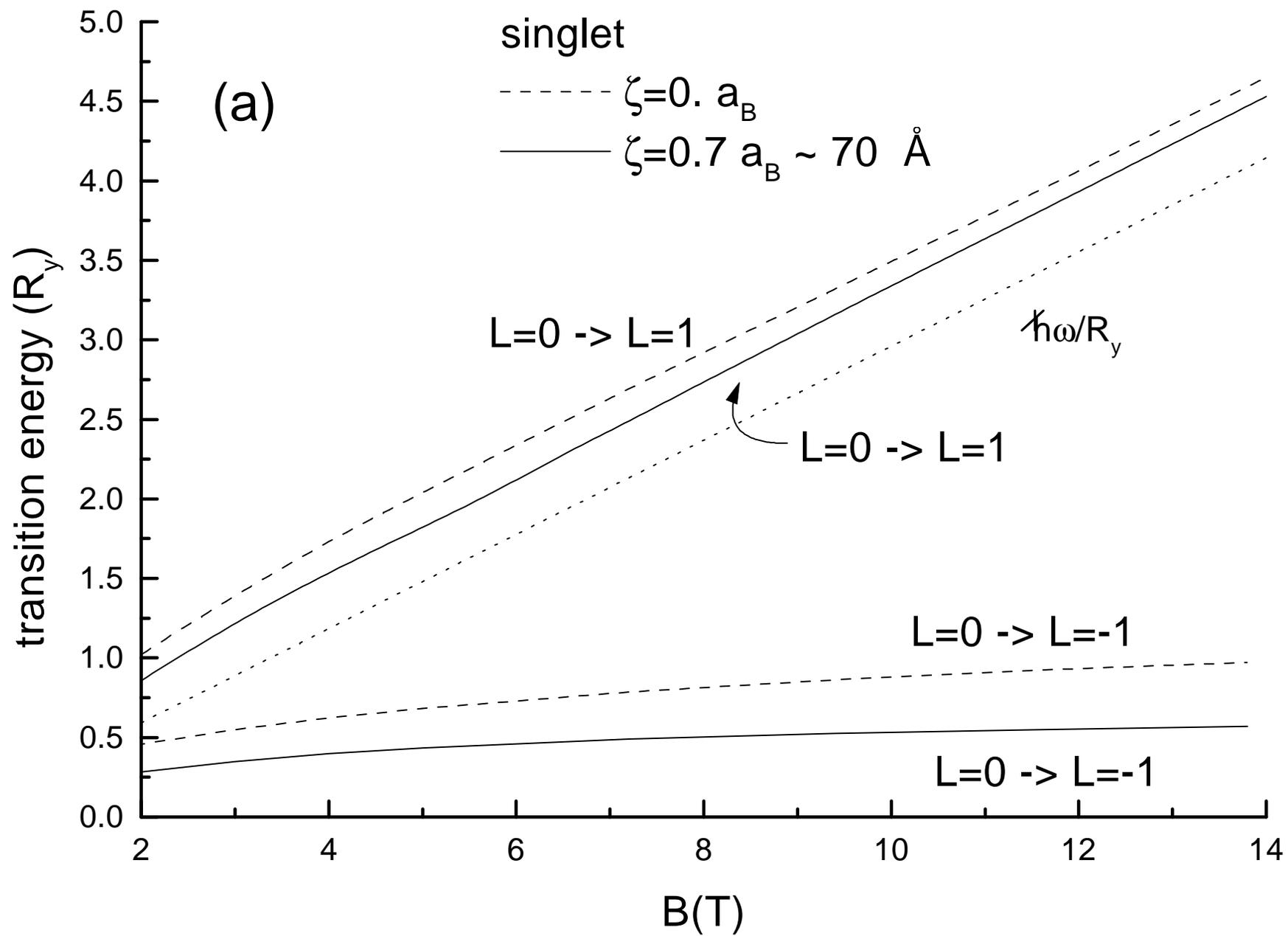
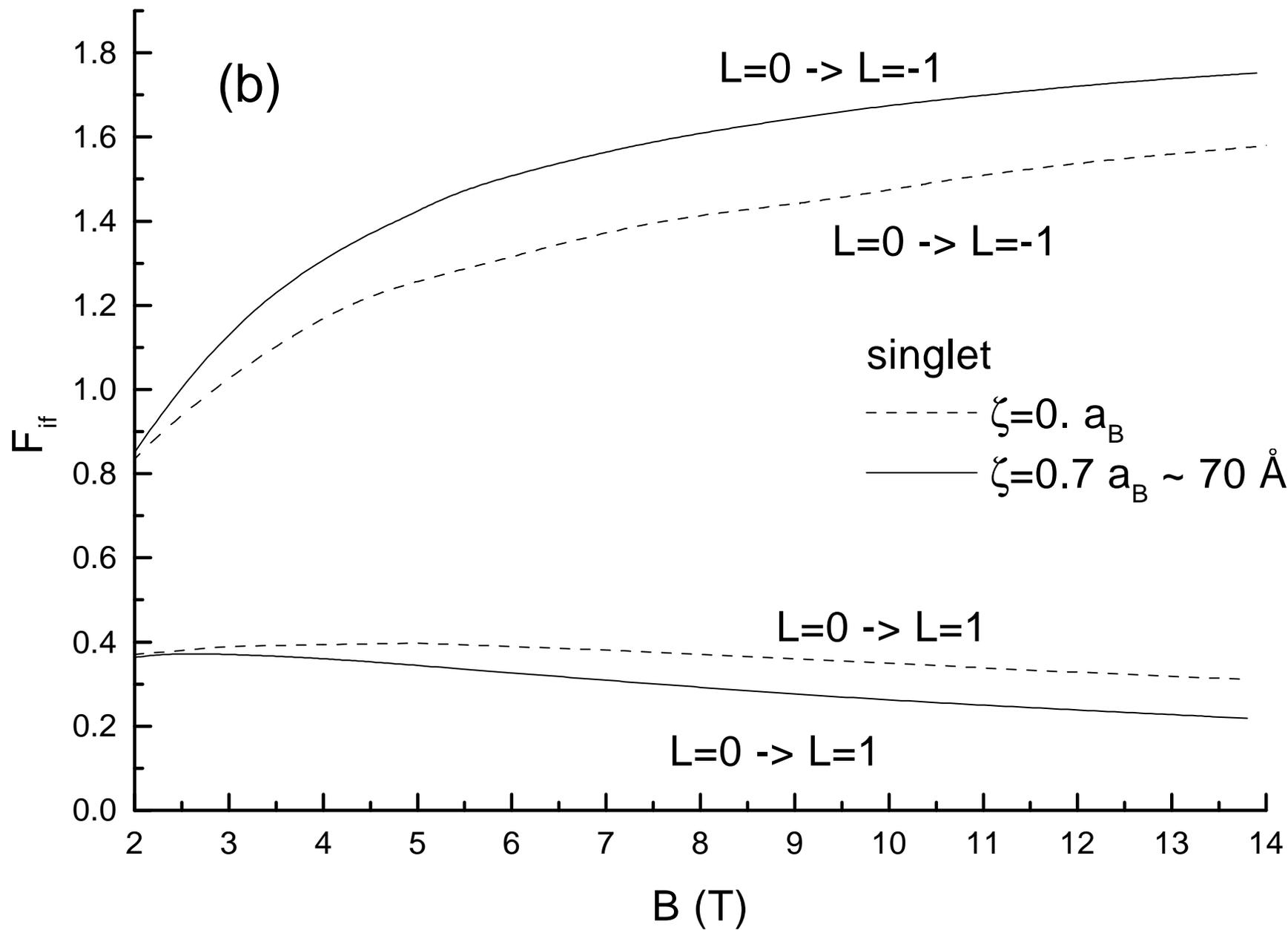
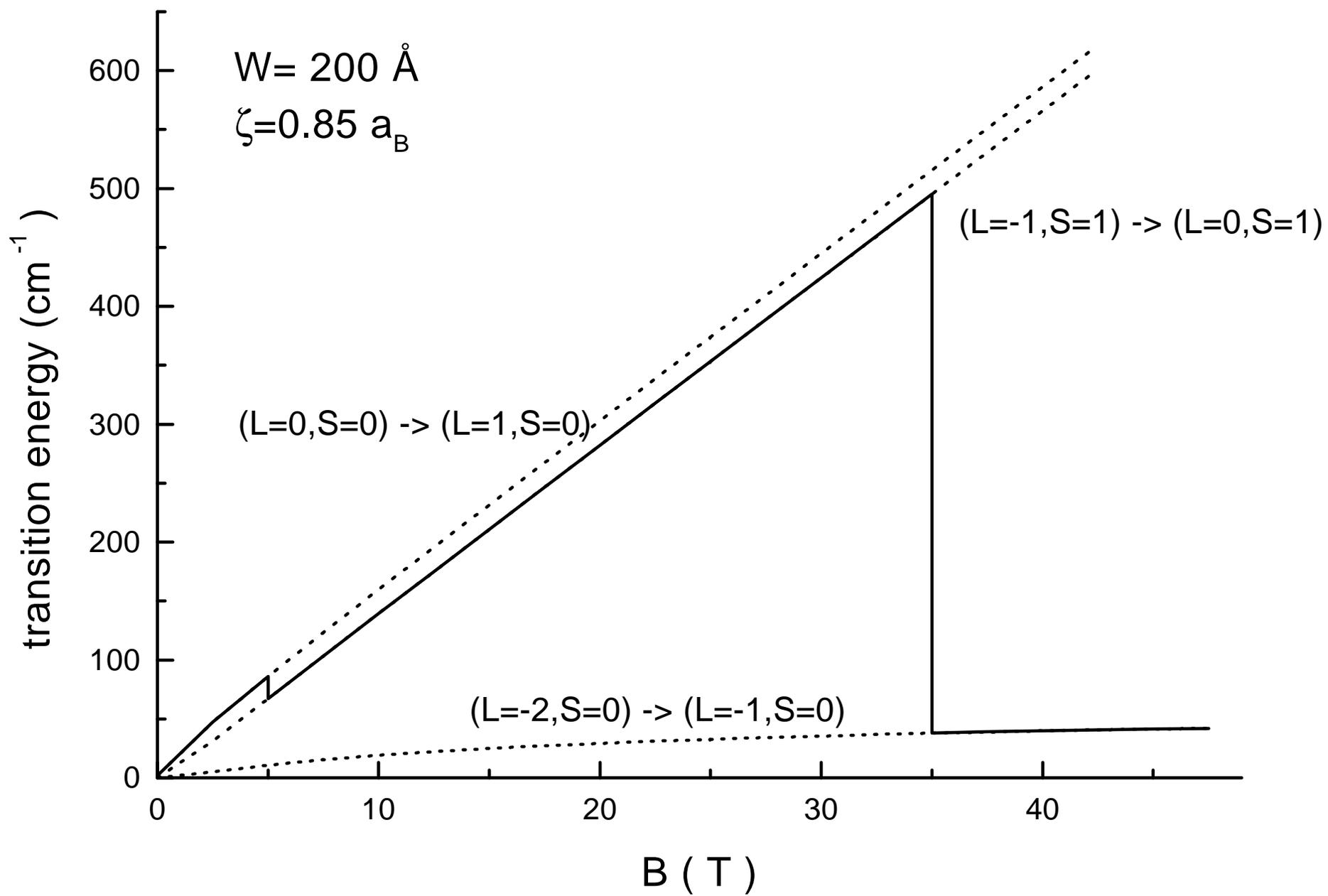
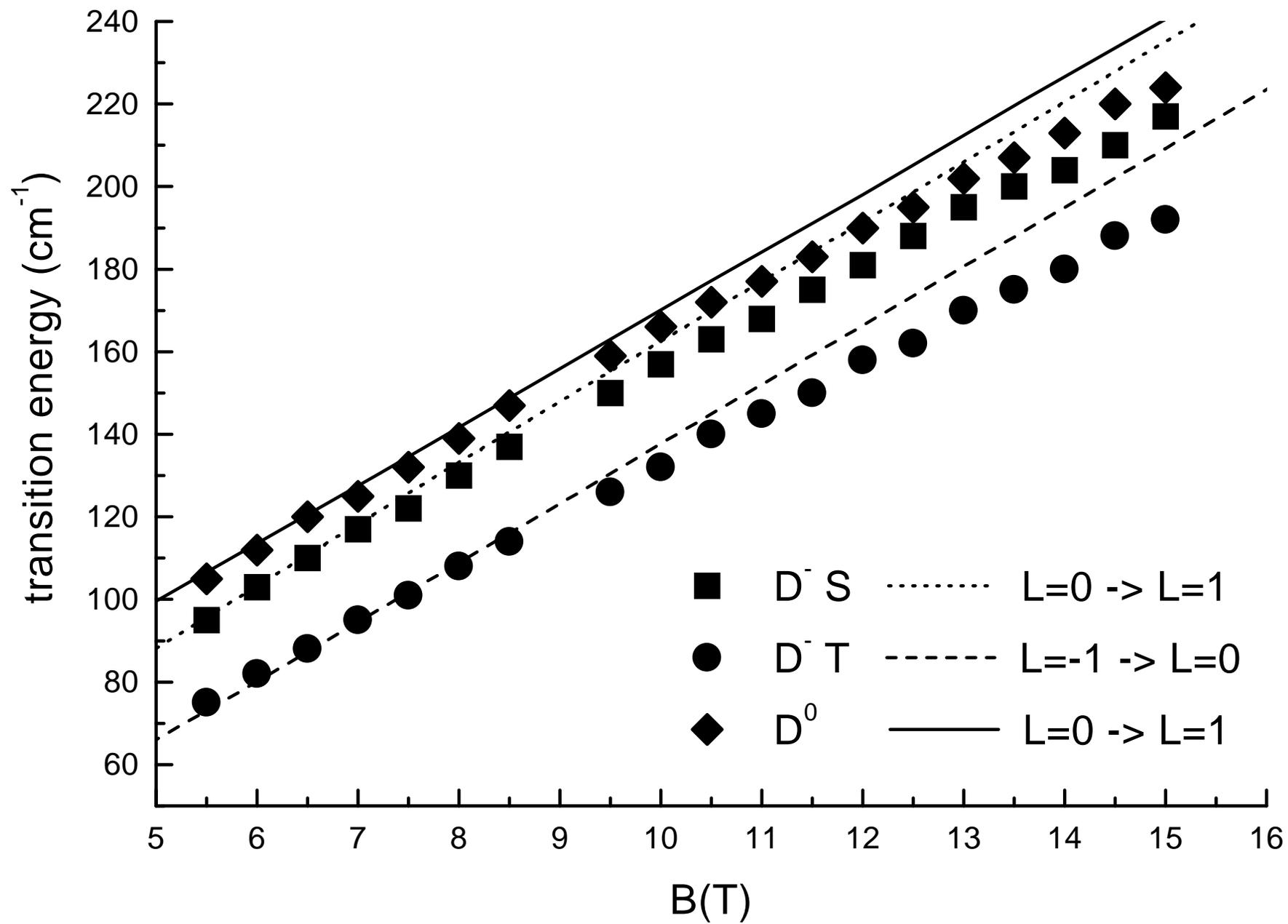
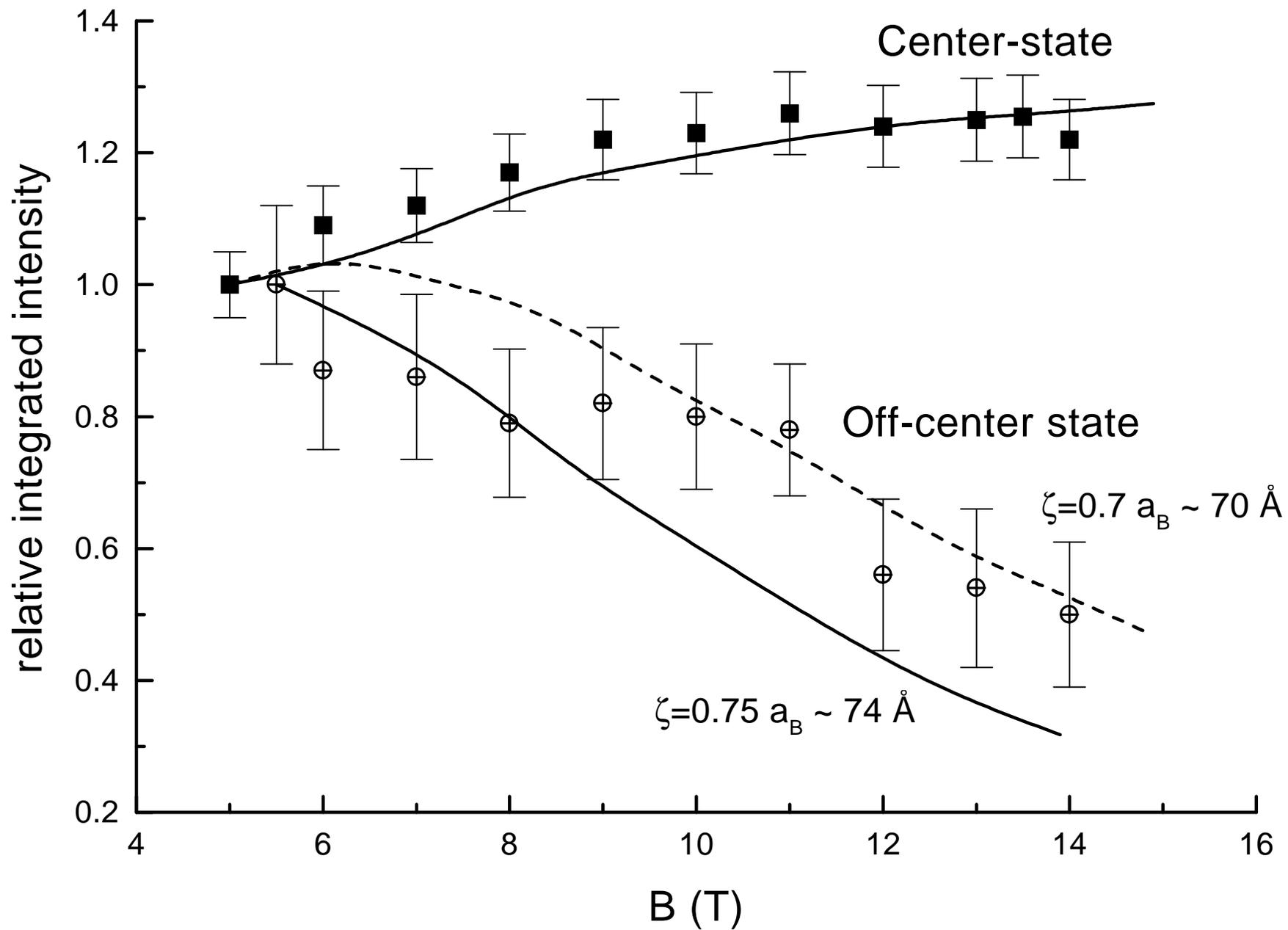
\end{document}